\documentclass[preprint]{aastex63}
\usepackage{ctable, bm, physics} 

\usepackage{amsmath}
\usepackage{multirow}

\newcommand\Rone{\textbf{\uppercase\expandafter{\romannumeral1}}\,}
\newcommand\Rtwo{\textbf{\uppercase\expandafter{\romannumeral2}}\,}
\definecolor{forestgreen}{rgb}{0.10, 0.50, 0.10}

\newcommand{\pt}[2]{\ensuremath{\dfrac{\partial\,#1}{\partial #2}}}
\newcommand{\ptn}[2]{\ensuremath{\dfrac{\partial #1}{\partial #2}}}
\newcommand{\dt}[2]{\ensuremath{\dfrac{\dd\,#1}{\dd #2}}}
\newcommand{\ha}[1]{\bm{#1}} 
\def\er{\ensuremath{\bm{e}_r}}
\def\et{\ensuremath{\bm{e}_\vartheta}}
\def\ep{\ensuremath{\bm{e}_\varphi}}
\def\ex{\ensuremath{\bm{e}_x}}
\def\ey{\ensuremath{\bm{e}_y}}
\def\ez{\ensuremath{\bm{e}_z}}
\def\br{\ensuremath{b_r}}
\def\bt{\ensuremath{b_\vartheta}}
\def\bp{\ensuremath{b_\varphi}}
\def\ur{\ensuremath{U_r}}
\def\ut{\ensuremath{U_\vartheta}}
\def\up{\ensuremath{U_\varphi}}

\def\Br{\ensuremath{B_r}}
\def\Bt{\ensuremath{B_\vartheta}}
\def\Bp{\ensuremath{B_\varphi}}

\def\utt{\ensuremath{U_{\vartheta_2}}}
\def\upt{\ensuremath{U_{\varphi_2}}}

\def\ett{\ensuremath{\ha{e_{\vartheta_2}}}}
\def\ept{\ensuremath{\ha{e_{\varphi_2}}}}

\definecolor{forestgreen}{rgb}{0.10, 0.50, 0.10}

\received{}
\revised{}
\accepted{}
\submitjournal{Science China Physics, Mechanics \& Astronomy}

\shorttitle{FastQSL}
\shortauthors{Chen et al.}

\begin{document}

\title{FastQSL 2: A Comprehensive Toolkit for Magnetic Connectivity Analysis}

\author[0000-0003-3060-0480]{Jun Chen}
  \affiliation{Purple Mountain Observatory, Chinese Academy of Sciences, Nanjing 210023, China}
  \affiliation{
Max-Planck-Institute for Solar System Research, Justus-von-Liebig-Weg 3, 37077 Göttingen,
Germany}
\correspondingauthor{Jun Chen}
\email{chenjun@pmo.ac.cn}

\author[0000-0001-6238-0721]{Thomas Wiegelmann}
  \affiliation{
Max-Planck-Institute for Solar System Research, Justus-von-Liebig-Weg 3, 37077 Göttingen,
Germany}

\author[0000-0003-4655-6939]{Li Feng}
  \affiliation{Purple Mountain Observatory, Chinese Academy of Sciences, Nanjing 210023, China}
\affiliation{
  School of Astronomy and Space Science, University of Science and Technology of China,
Hefei 230026, China}
\correspondingauthor{Li Feng}
\email{lfeng@pmo.ac.cn}

\author[0000-0002-7018-6862]{Chaowei Jiang}
\affiliation{State Key Laboratory of Solar Activity and Space Weather, School of Aerospace Science, Harbin Institute of Technology, Shenzhen 518055, China}

\author[0000-0003-4618-4979]{Rui Liu}
\affiliation{CAS Key Laboratory of Geospace Environment,
	 School of Earth and Space Sciences, \\
	 University of Science and Technology of China,
	  Hefei, Anhui 230026, China}
\affiliation{CAS Center for the Excellence in Comparative Planetology, 
    \\University of Science and Technology of China,  Hefei, Anhui 230026, China}



\begin{abstract}
We present a new version of FastQSL for locating quasi-separatrix layers (QSLs) --- regions characterized by strong magnetic connectivity gradients, preferential current buildup, and subsequent magnetic reconnection.
This version now supports spherical coordinates, utilizing a second spherical coordinate system for tracing magnetic field lines around the polar regions. This approach completely resolves the singularity problem at the two poles. Furthermore, our code accommodates arbitrary mesh shapes for output, can provide both magnetic field and electric current density on the mesh, and can save the traced magnetic field lines. 
We suggest using $Q_\mathrm{local}$ calculated through a localized mapping to locate (quasi-)separators.
By quickly and accurately outputting the footpoint coordinates of magnetic field lines, FastQSL can be used to derive the two key parameters used for modeling solar wind speed and slip-squashing factors for the case of zero boundary flow. Compared with the first version, FastQSL 2 achieves significant improvements in terms of application scope.
\end{abstract}

\keywords{Solar corona, Magnetic topology, Open source software}


\section{Introduction}
Magnetic field lines can bifurcate at some sites. A converging plasma flow around such sites can efficiently cancel the anti-parallel component of the magnetic field through electrical resistance \citep{Longcope1994a, Longcope1994b}, a process known as magnetic reconnection. 
Magnetic field lines can be classified into distinct topologies, separated by the zero-thickness magnetic surfaces that originate at bifurcation sites.
Such a surface is called a separatrix by \cite{Longcope1994a, Longcope1994b}.
A separator, termed the intersection line of separatrices, is the site where magnetic field lines bifurcate.
A localized region where the change in magnetic connectivity is steep but finite is designated a quasi-separator.
The field lines originating from a quasi-separator form QSLs \citep{priest1995three, demoulin1996three, Demoulin1997,Priest2000,Pontin2011}, which have a non-zero thickness.

Separators and quasi-separators host favorable conditions for three-dimensional magnetic reconnection. Chromospheric flare ribbons are frequently observed to align with the footprints of separatrices or QSLs \citep{priest1995three,demoulin1996three,Demoulin1997}. \cite{Chen2020} found the EUV late phase \citep{Woods2011} of circular-ribbon flares is mostly emitted from the region around a plate-shaped QSL. 
A ballooning instability in the
near-Earth magnetotail can induce QSLs \citep{Zhu2019}. \cite{Gekelman2020} used QSLs to locate the position of three‑dimensional reconnection in laboratory plasma.

In \cite{Titov2002} and \cite{Titov2007},
such magnetic connectivity can be defined by a field-line mapping. 
For a field line launched from the point $\bm{R}_u$  on a surface $S_u$ coordinated with $(u_1, u_2)$, its target point $\bm{R}_w$ on a surface $S_w$ can be described by the coordinates $(w_1, w_2)$. Based on the mapping of $(u_1, u_2) \rightarrow (w_1,w_2)$, the change of magnetic connectivity at $\bm{R}_u$ 
can be quantified as
\begin{align}
  Q|_{\bm{R}_u}=
  \left.
  \left(\sum_{i,j,k,l}
  \pt{w_i}{u_k}\,{g}_{w,ij}\,\pt{w_j}{u_l}\,g_u^{lk}\right)
  \,
  \right/
  \left(
  \sqrt{\frac{g_w}{g_u}}\,\left|\ptn{(w_1,w_2)}{(u_1,u_2)}\right|
  \right),
  \label{eq:q}
\end{align}
where 
\begin{align}
g_{w,ij} = \ptn{\bm{R}}{w_i}\cdot\ptn{\bm{R}}{w_j}
\end{align}
is the component of covariant metric tensor on $S_w$, 
and $g_{u,ij}$ can be defined in the same way;
$g_u =\mathrm{det}[g_{u,ij}]$, and $g_w=\mathrm{det}[g_{w,ij}]$; 
$g_u^{lk}$ is the component of contravariant metric tensor
on $S_u$; 
the Jacobian determinant 
\begin{align}
\ptn{(w_1,w_2)}{(u_1,u_2)}=\ptn{w_1}{u_1} \ptn{w_2}{u_2}-\ptn{w_2}{u_1} \ptn{w_1}{u_2}.
\end{align}
\cite{Titov2002} and \cite{Titov2007} proved the identity $Q|_{\bm{R}_u}=Q|_{\bm{R}_w}$, thus,
\cite{Aulanier2005} expanded the definition of $Q$ into 3D space using the invariant $Q$-value along a field line, i.e.
\begin{align}
\bm{B} \cdot \nabla Q=0.
\label{eq:BcQ}
\end{align}
Separatrices can be characterized by an infinite $Q$-value, and QSLs are characterized by a $Q$-value that significantly exceeds its theoretical minimum of 2.

To mitigate numerical error, an identity derived from $\nabla \cdot \bm{B}=0$ is usually applied to the denominator of Equation \eqref{eq:q},
\begin{align}
\sqrt{\frac{g_w}{g_u}}\,\left|\ptn{(w_1,w_2)}{(u_1,u_2)}\right|=
\left|\frac{B_{n,u}}{B_{n,w}}\right|,
\label{eq:Delta}
\end{align}
where $B_{n,u}$ and $B_{n,w}$ are the normal components of magnetic field on $S_u$ and $S_w$, respectively. 
For numerical fields, $\nabla \cdot \bm{B}=0$ is not always strictly enforced, potentially introducing errors into the $Q$-values. Nevertheless, the locations of QSLs, identified by high-$Q$, are still generally reliable. In fact, the first approach for locating QSLs, which used the high value of the square root of the numerator in Equation \eqref{eq:q}, denoted $N$ in \cite{priest1995three}, can still provide correct locations. Since $N|_{\bm{R}_u} \ne N|_{\bm{R}_w}$ when $|B_{n,u}| \ne |B_{n,w}|$, 
this discrepancy leads to an inconsistent measure of connectivity change along the same field line. 
This is the reason why $Q$ is now the standard parameter for locating QSLs.

A magnetic field line is generally not perpendicular to $S_{u}$ and $S_{w}$ at its intersection points $\bm{R}_u$ and $\bm{R}_w$,  respectively.
This inclination introduces a projection effect to Equation \eqref{eq:q}.
The correction involves introducing two local surfaces: 
surface $S_{u,\perp}$ through $\bm{R}_u$  and
the surface $S_{w,\perp}$ through $\bm{R}_w$, each oriented perpendicular to the field line. 
These surfaces are equipped with coordinates $(u_{1,\perp},\,u_{2,\perp})$ and  $(w_{1,\perp},\,w_{2,\perp})$, respectively.
Applying their coordinate mapping through magnetic field lines, $(u_{1,\perp},\,u_{2,\perp})\rightarrow (w_{1,\perp},\,w_{2,\perp})$,  into
Equation \eqref{eq:q}
will yield $Q_\perp$ at $\bm{R}_u$ and $\bm{R}_w$. 
To obtain $Q_\perp$ at every point on the original surfaces $S_{u}$ and $S_{w}$, 
the local perpendicular surfaces 
$S_{u,\perp}$ and $S_{w,\perp}$ are determined as the planes orthogonal to the field line at its intersection with  $S_{u}$ and $S_{w}$, respectively.
$Q_\perp$ provides a more precise quantification of the connectivity change within a volume than $Q$.

$Q$ and $Q_\perp$ can be provided by several open-source codes, 
such as K-QSL
\footnote{\url{https://github.com/Kai-E-Yang/QSL}}
\citep{K-QSL}, 
QSL-Squasher 
\footnote{\url{https://bitbucket.org/tassev/qsl_squasher/}}
\citep{tassev2017qsl}, 
UFiT 
\footnote{\url{https://github.com/Valentin-Aslanyan/UFiT}}
\citep{Aslanyan_2024},
MapFL \footnote{\url{https://github.com/predsci/MapFL}}, 
and FastQSL
\citep{FastQSL}.
FastQSL achieves very high performance by
giving millions of $Q$ values per second \citep{FastQSL}.
However, FastQSL did not support spherical coordinates in the first version (hereafter FastQSL 1) . 

In this work, we present a development of FastQSL, termed FastQSL 2, 
which maintains the high performance of FastQSL 1 and is also available online\footnote{\url{https://github.com/el2718/FastQSL2}}.
In Section \ref{s:Spherical}, 
we detail the analytical forms of the squashing factor $Q$ in spherical coordinates and validate the capability of computing $Q$ on a spherical grid.
In Section \ref{s:Functionalities}, we show features and products of FastQSL 2,
and show how to use these products to derive other quantities.
In Section \ref{s:summary}, we discuss and summarize the developments.




\section{Extension for Spherical Coordinates} \label{s:Spherical}

\begin{figure}[ht]
    \centering
    \includegraphics[width=\linewidth]{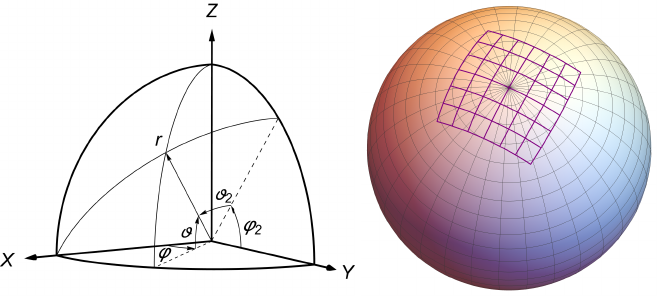}
    \caption{Left: two-spherical-coordinate system.
    Right: the mesh around a polar region. 
    The grids of $\varphi, \vartheta$
    are shown by the crosses of the gray mesh;
    the grids of $\varphi_2, \vartheta_2$
    are shown by  the crosses of the purple mesh.
    }
    \label{fig:coordinates}
\end{figure}

As shown by the left panel of Figure \ref{fig:coordinates}, we denote longitude (in radians), latitude (in radians), and radius by $\varphi$, $\vartheta$, and $r$, respectively.
The relations between $\{\varphi,\,\vartheta ,\,r\}$ and $\{x,\,y,\,z\}$ are
\begin{align}
x&=r\,\cos \vartheta\,\cos \varphi,\label{eq:x_spherical}\\
y&=r\,\cos \vartheta\,\sin \varphi,\label{eq:y_spherical}\\
z&=r\,\sin \vartheta.              \label{eq:z_spherical}
\end{align}
The corresponding unit elementary vectors are
\begin{align}
\ep&=-\sin \varphi\, \ex+\cos \varphi\,\ey,\label{eq:ep}\\
\et&=-\sin \vartheta \,\cos \varphi\, \ex-\sin\vartheta\,\sin\varphi\, \ey+\,\cos\vartheta\, \ez,\label{eq:et}\\
\er&=\cos \vartheta\,\cos \varphi\, \ex+\cos\vartheta\,\sin \varphi\, \ey+\sin\vartheta\, \ez. \label{eq:er}
\end{align} 
In this spherical coordinate system,
\begin{align}
\nabla \times \bm{B}&=
\frac{\ep}{r}\left(\pt{\Br}{\vartheta}-\pt{(r\,\Bt)}{r}\right)+ 
\frac{\et}{r}\left(\pt{(r\,\Bp)}{r}-\frac{1}{\cos \vartheta} \pt{\Br}{\varphi}\right)+
\frac{\er}{r\,\cos \vartheta}\left(\pt{\Bt}{\varphi}-\pt{(\cos \vartheta\,\Bp)}{\vartheta}\right).
\end{align}
A magnetic field line can be traced by
\begin{align}
    \frac{ \dd \{\varphi,\, \vartheta,\,r\}}{\dd s}=
    \left\{\frac{\bp}{r\, \cos \vartheta}, \,\frac{\bt}{r},\,\br \right\},
    \label{eq:flb}
\end{align}
where $\bp=\Bp/|B|$, $\bt=\Bt/|B|$, $\br=\Br/|B|$, and $s$ is the arc length coordinate of a field line.

\subsection{The Spherical Form of Calculating $Q$} \label{ss:Qmethod}

The distribution of $Q$ on a cross section $S_h$ can be defined by Equation \eqref{eq:BcQ}.
If we launch a field line from $\bm{R}_h$ on $S_h$, the tracing can yield two target points, $\bm{R}_u$ on $S_u$ and $\bm{R}_w$ on $S_w$. $S_h$ is equipped with coordinates $(h_1, h_2)$, 
then the field line tracings launched from $S_h$ make two mappings $(h_1, h_2)\rightarrow (u_1, u_2)$ and $(h_1, h_2)\rightarrow (w_1, w_2)$, 
then the mapping $(u_1, u_2)\rightarrow (w_1, w_2)$ can be derived, and we can define $Q|_{\bm{R}_h}$ on $S_h$
from Equation \eqref{eq:q}. 

Method 3 of \cite{Pariat2012} (hereafter Method \Rone) rewrites Equation \eqref{eq:q} by including the auxiliary coordinates $(h_1, h_2)$, and calculates the partial derivatives in Equation by finite differences, using 
the coordinates of neighboring target points. 
The corresponding formulas (17-20) in \cite{Pariat2012}
are written in Cartesian coordinates; 
the expressions in spherical coordinates are derived in Appendix \ref{ap:method1}.

An alternative way to calculate $Q$ has been proposed by \cite{Scott2017}  (hereafter Method \Rtwo). 
For two vectors $\bm{U}$ and $\bm{V}$ obeying
\begin{align}
\lambda\, \bm{B} \cdot \nabla \bm{U} &=\bm{U} \cdot \nabla (\lambda\, \bm{B})
\label{eq:bu}\\
\lambda\,\bm{B} \cdot \nabla \bm{V} &=\bm{V} \cdot \nabla (\lambda\, \bm{B})
\end{align}
along a field line,
where $\lambda$ can be any positive function in the space,
\cite{Scott2017} proved 
\begin{align}
Q = \frac{ \Tilde{\bm{U}}_u^2\, \Tilde{ \bm{V}}_w^2 +
\Tilde{\bm{U}}_w^2\, \Tilde{ \bm{V}}_u^2
-
2\,(\Tilde{\bm{U}}_u\cdot\Tilde{ \bm{V}}_u)\,
(  \Tilde{\bm{U}}_w\cdot \Tilde{\bm{V}}_w)}
{(B_{n,h})^2/ |B_{n,u}\, B_{n,w}|},
\label{eq:q_scott}
\end{align}
where
\begin{align}
\Tilde{\bm{U}}_u= \bm{U}-{\left.
\frac{ \bm{U}\cdot \bm{n} }{ \bm{B}\cdot \bm{n} }\,\bm{B} \right|}_{S_u},
\end{align}
and $\bm{n}$ is the normal direction vector of $S_u$;
$B_{n,h}$ is the normal component of magnetic field on $S_h$.
The forms are similar for $\Tilde{\bm{U}}_w,\,\Tilde{\bm{V}}_u,\,\Tilde{\bm{V}}_w$.
$Q_\perp$ can be directly achieved by setting $\bm{n}=\bm{B}$.
\cite{Aslanyan_2024} chooses $\lambda=1$ and gives Equations (18-20) in their work. 
In FastQSL, we parameterize a field line with the arc length coordinate $s$, so we set $\lambda = 1/|B|$. This yields $\dfrac{\bm{B}}{|B|}\cdot \nabla = \dt{}{s}$, and Equation \eqref{eq:bu} then becomes
\begin{align}
  \dt{\up}{s}= &
  \dfrac{\up}{r\,\cos \vartheta}\, 
      \left(\pt{\bp}{\varphi}-\bt\,\sin\vartheta+\br\,\cos\vartheta\right)+
  \frac{\ut}{r}\left(\frac{\sin\vartheta}{\cos\vartheta}\,\bp+\pt{\bp}{\vartheta}\right)+
  \ur\,\left(\pt{\bp}{r}-\frac{\bp}{r}\right),
  \label{eq:dup}
\\
  \dt{\ut}{s}=&
\dfrac{\up}{r\,\cos \vartheta}\,
\pt{\bt}{\varphi}+
\dfrac{\ut}{r}\,\left(\pt{\bt}{\vartheta}+ \br \right)+
\ur\,\left(\pt{\bt}{r}-\frac{\bt}{r} \right),
\label{eq:dut}
\\
\dt{\ur}{s} =&
\dfrac{\up}{r\,\cos \vartheta}\,\pt{\br}{\varphi}+
\dfrac{\ut}{r}\,\pt{\br}{\vartheta}+\ur\,\pt{\br}{r}.
\label{eq:dur}
\end{align}
The vector $\bm{V}$ obeys identical expressions.

The mapping of $\bm{U}$ and $\bm{V}$ avoids the coordinate difference error associated with Method \Rone. However, since the output mesh is discretized, Method \Rtwo could fail to capture thin QSLs or even zero-thickness separatrix surfaces, where the $Q$-distribution can exhibit sharp gradients \citep{FastQSL}.
MapFL only supports Method \Rone;
K-QSL, QSL-Squasher, UFiT only support Method \Rtwo;
FastQSL continues to support both methods, now also in spherical coordinates.
For the default setting which we recommend, most values of $Q$ are given by Method \Rone;
for some grids that 
do not allow the calculation of the coordinate difference, the values of $Q$ are filled using Method \Rtwo. In FastQSL, 
$Q_\perp$ can only be achieved by Method \Rtwo.

\subsection{A Secondary Coordinate System for the Polar Regions}

Equation \eqref{eq:flb} exhibits a singularity at the poles
where the latitude $\vartheta=\pm\, \pi/2$. We solve this problem by
introducing a second spherical coordinate system $\{\varphi_2,\,\vartheta_2,\,r\}$ 
that obeys
\begin{align}
y&=r\,\cos \vartheta_2\,\cos \varphi_2,\label{eq:y_spherical2}\\
z&=r\,\cos \vartheta_2\,\sin \varphi_2,\label{eq:z_spherical2}\\
x&=r\,\sin \vartheta_2.                \label{eq:x_spherical2}
\end{align}
As shown by the right panel of Figure \ref{fig:coordinates}, 
if the input grid includes the poles,
FastQSL will generate a second mesh around the polar region of the first mesh, 
and will interpolate the magnetic field $B_{\varphi_2},\, B_{\vartheta_2},\, B_r$ on the second mesh.
During field line tracing via Equation \eqref{eq:flb}, transformations of  coordinates, vectors, and directional derivatives are implemented when positions cross the polar boundary $\abs{\vartheta} = \vartheta_\mathrm{polar}$ 
 (see Appendix \ref{ap:transformation}), and $\vartheta_\mathrm{polar}$ is set to $3\,\pi/8$ in our code.

\subsection{Validation of the quality of $Q$} \label{s:Validation}
\begin{figure}[ht]
    \centering
    \includegraphics[width=\linewidth]{fig/figure2.pdf}
    \caption{
    Left: $B_r$ at $r=1.04~R_\mathrm{sun}$ with selected magnetic field lines from both the inserted flux rope and the background field overlaid; the distribution of the twist number is displayed on a planar cross-section.
Right: the distributions of $Q$ on the surface of $r=1.04~R_\mathrm{sun}$, the same planar cross-section, and the surface of a hollowed-out dome;  
the purple region marks one segment of a separator detected with $Q_\mathrm{local} > 1000$ (with $r_\mathrm{local}=0.02~R_\mathrm{sun}$).
}
\label{fig:FMG-Quadrupole2}
\end{figure}
As illustrated in Figure \ref{fig:FMG-Quadrupole2}, we show the magnetic topology of an extrapolated field derived from a solar synoptic magnetic map observed by the Full-disk vector MagnetoGraph (FMG; \cite{Deng2019FMG}) on board 
the Advanced Space-based Solar Observatory (ASO-S; \cite{Gan2023ASOS}).
The working spectral line of FMG is Fe  \uppercase\expandafter{\romannumeral1} 532.4~nm; the system-wide spatial resolution is $1.5^{\prime \prime}$; the pixel resolution is $0.5^{\prime \prime}$.
The magnetic synoptic map spans the period from UTC 11:49:55 on May 6, 2024 to UTC 16:59:31 on June 2, 2024 (Carrington rotation 2284). It is constructed from the line-of-sight magnetic field component measured at a strip centered at the central meridian at six-hour intervals. The grid size of the magnetic synoptic map is  $3660\times1801$.
The extrapolation is performed using the Potential Field Source Surface (PFSS) model \citep{PFSS1984,PFSS1992ApJ} with  POT3D \footnote{\url{https://github.com/predsci/POT3D}} \citep{pot3d}. The location of the source surface is $r=2.5~R_\mathrm{sun}$, where $R_\mathrm{sun}$ is the solar radius. 
The $B_r$ for PFSS model is directly set to the line-of-sight magnetic field component; due to the observational limitations, the $B_r$ at the high latitude region is unreliable. This extrapolation is only used for illustrating the result of FastQSL, so such an approximation to $B_r$ should be acceptable. 
To present large-scale connectivities, the first two layers from the bottom 
($r=R_\mathrm{sun}$) of the original extrapolated field are removed, as they can introduce many localized, small-scale connectivity features; 
therefore the first layer of $r$-direction is changed to $r=1.04~R_\mathrm{sun}$.
Finally, the grid size of the extrapolated field is $361\times181\times74$.
In order to demonstrate the field line tracing capability around the poles, the field of Quadrupole2 described in \cite{chen2023model} is embedded within the north polar region.

\begin{figure}[ht]
    \centering
    \includegraphics[width=\linewidth]{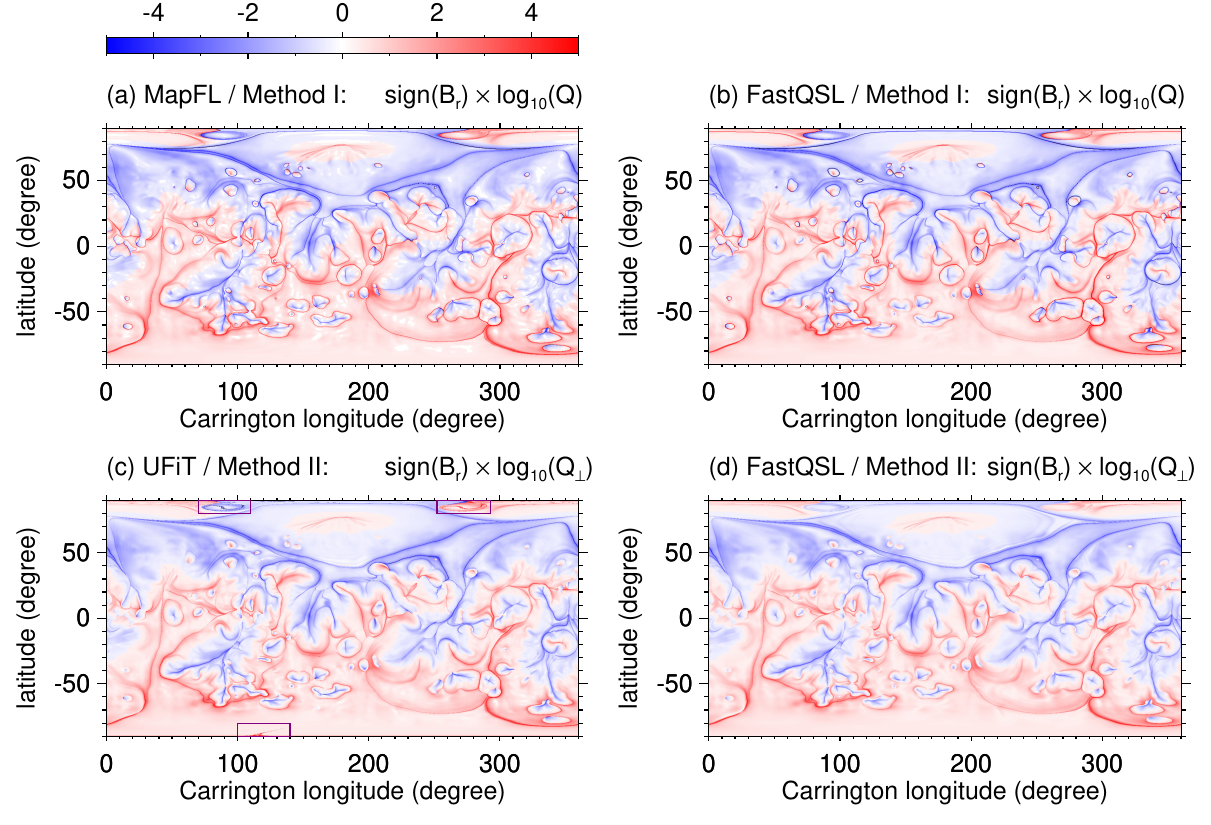}
    \caption{The $Q$-map at $r=1.04~R_\mathrm{sun}$ of the same field used in Figure \ref{fig:FMG-Quadrupole2}.
    (a) The map of $Q$ computed with MapFL using Method \Rone.
    (b) The map of $Q$ computed with FastQSL using Method \Rone.
    (c) The map of $Q_\perp$ computed with UFiT using Method \Rtwo.
    (d) The map of $Q_\perp$ computed with FastQSL using Method \Rtwo.
}
    \label{fig:compare_code}
\end{figure}
We validate the quality of $Q$ from FastQSL by comparing the results from MapFL and UFiT. 
MapFL traces a field line by Euler integral; the step size is scaled according to the field line's radius of curvature, and the parameter \texttt{ds\%over\_rc} for this scheme is set to 0.02 in Figure \ref{fig:compare_code} (a).
UFiT can trace a field line by the classic \texttt{RK4}, and the step size is set to 0.008 in Figure \ref{fig:compare_code} (c).
For FastQSL, we use the 3/8-rule \texttt{RK4} \citep{kuttaRK4} to trace field lines;
the integration step size is set to 0.9 for Figure \ref{fig:compare_code} (b);
the integration step size is set to 1.9 for Figure \ref{fig:compare_code} (d).
Please note that the step sizes in UFiT and FastQSL use different units.
For each code, the parameters are chosen to be near the threshold that yields a marginal ground truth image \citep{FastQSL}; i.e., the image obtained with a smaller parameter value shows no significant difference from the marginal ground truth image.
The $Q$-map in Figure \ref{fig:compare_code} has a grid size of $1441 \times 721$. The computation times were as follows: MapFL took 6.09 seconds for Figure \ref{fig:compare_code} (a), UFiT took 3.71 seconds for Figure \ref{fig:compare_code} (c), and FastQSL took 1.59 seconds for Figure \ref{fig:compare_code} (b) and 1.15 seconds for Figure \ref{fig:compare_code} (d). All computations were performed on an Apple M3 Pro chip using gfortran 15.2.0.
The positions of QSLs presented in Figure \ref{fig:compare_code} (a) and Figure \ref{fig:compare_code} (b) are identical. The low-$Q$ regions in Figure \ref{fig:compare_code} (b) are slightly smoother, which is not very important for QSL-detection.
Comparing Figure \ref{fig:compare_code} (c) and Figure \ref{fig:compare_code} (d), the $Q$-maps in most areas are indistinguishable. However, for the polar regions, some pseudo-structures appear in Figure \ref{fig:compare_code} (c), as marked by the three purple boxes.

\section{Functionalities} \label{s:Functionalities}
\subsection{Inherited and Enhanced Features from FastQSL 1} \label{s:inherite}
FastQSL 1 has two sets of code.
The first set 
\footnote{\url{https://github.com/el2718/FastQSL}} is directly extended from the code published with \cite{Liu2016apj}. Its interface for input and output (IO) runs on IDL \footnote{\url{https://www.nv5geospatialsoftware.com/Products/IDL}}, and its computation core is written with Fortran
\footnote{\url{https://fortran-lang.org/}}.
The second set 
\footnote{\url{https://github.com/peijin94/FastQSL}}
is accelerated by an NVIDIA GPU. Its interface for IO runs on Python
\footnote{\url{https://www.python.org/}}
, and its computation core is written with CUDA/C
\footnote{\url{https://docs.nvidia.com/cuda/cuda-c-programming-guide/index.html}}
.

FastQSL 2 inherits most features from the first set of FastQSL 1, the main inheritances (with some extensions) are: \footnote{The author of the second set of FastQSL 1, Peijin Zhang, has moved to USA. He doesn't have time for the extension with GPU. 
Since we believe the first set of FastQSL 1 is fast enough for $Q$-computation, currently we have no plans to accelerate FastQSL 2 by a GPU.}
\begin{enumerate}
\item The interface for IO runs on IDL; now the interface for IO can also run on Python. The computation core is written with Fortran. The operating system can be any of Linux, MacOS, and Windows.
\item Supporting a Cartesian, uniform or stretched grid. Now a spherical grid is also supported. 
For a Cartesian grid, periodic boundaries can now be specified using the keywords "xperiod", "yperiod", and "zperiod".
For example, invoking "yperiod" with $B_x$, $B_y$, $B_z$ data that are invariant in the $y$-direction reduces the effective dimensionality to 2D, rendering FastQSL capable of tracing field lines in a 2D space.
\item Trilinear interpolation (Equations (13--15) in \cite{FastQSL}) is applied to interpolate magnetic field, and the expressions are similar on a spherical mesh.
\item Tracing field lines with a step-fixed scheme of 3/8-rule \texttt{RK4} or
a step-size adaptive scheme of \texttt{RKF45}
\citep{fehlberg1969low}.
\item The coordinates of a terminated point can exactly lie on a boundary with the treatment described in Appendix \ref{ap:boundary_trim}.
\item Supporting both Method \Rone and Method \Rtwo to compute $Q$.
\item Producing $Q$ and other products (see Section \ref{s:products}).
\end{enumerate}

\subsection{Products} \label{s:products}

FastQSL can provide results within a domain of any shape.
Any cross section with one fixed coordinate (e.g. $\vartheta=\vartheta_0$)
is easy to set at the input of FastQSL.
Sometimes we want to examine the magnetic topology on a flat cross-section perpendicular to the photosphere, while the normal direction of the cross-section may be neither longitudinal nor latitudinal.
This surface is also easy to configure by invoking the keyword 'csFlag', 
using the expression in
Appendix \ref{ap:great_circle}. The distributions of $Q$ and the twist number on such a representative planar cross-section are displayed in Figure \ref{fig:FMG-Quadrupole2}.
In FastQSL 1, the output grid is described by specifying the lower and upper bounds of three coordinates and the grid spacing.
In FastQSL 2, the keyword 'seed' offers an alternative way to describe the output grid, as it can define a domain of any shape and dimension from 0 to 3.
For example, the right panel of Figure \ref{fig:FMG-Quadrupole2} presents the distribution of $Q$ on a hollowed-out dome surface computed with Method \Rone.

\begin{figure}[ht]
    \centering
    \includegraphics[width=\linewidth]{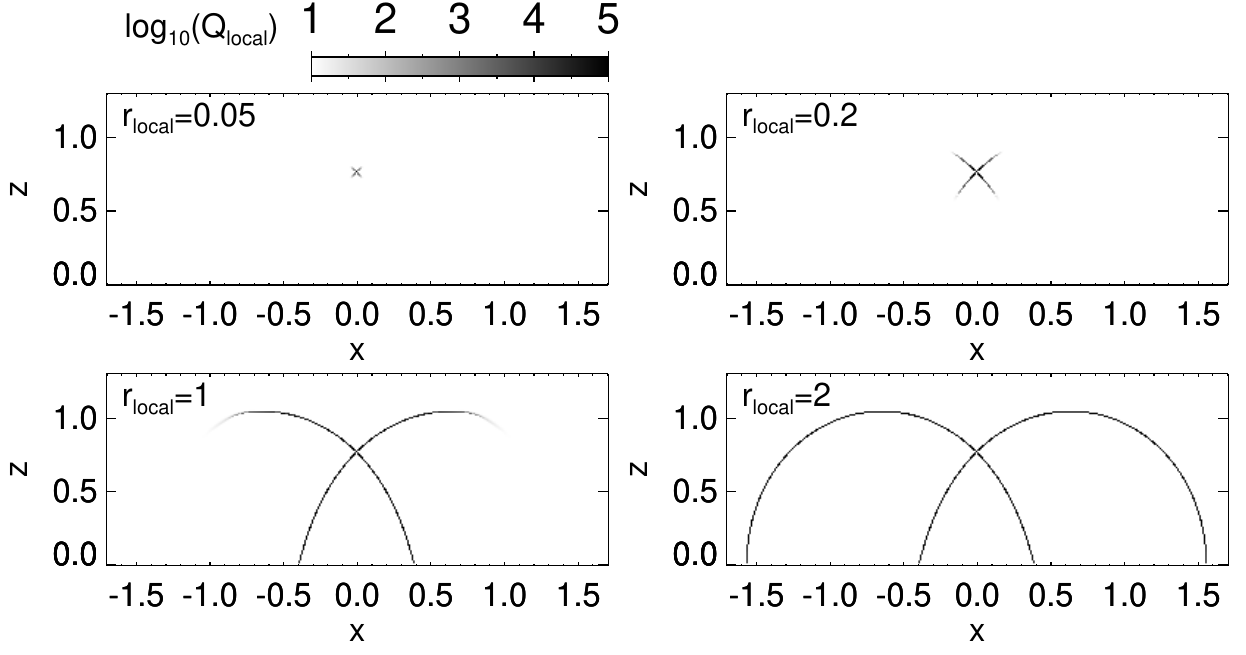}
    \caption{
    The distributions of $Q_\mathrm{local}$ for $r_\mathrm{local}=0.05,\, 0.2,\, 1,\, 2$ in the plane of $y=0$ with $\bm{B}_\mathrm{quadrupole}$. }
    \label{fig:qlocal}
\end{figure}

FastQSL can provide the following products on the output grid:
\begin{enumerate}
\item The path of every magnetic field line launched from the output grid, and $\bm{B}$ and $\nabla\times\bm{B}$ along the path.
For example, the field lines in Figure \ref{fig:FMG-Quadrupole2} are traced with FastQSL from 5 seed points.
\item
The coordinates of the two end points of every field line. Such high-accuracy coordinate mapping can be used to derive some other useful products related to the mapping, and two examples are shown in Section \ref{s:derived_products}.
\item
The type marks of 
the end point of every field line defined in Table \ref{tb:target}.
A mark of 0 may be encountered when tracing an extremely long field line. To prevent this issue in some special fields, the maximum steps of tracing can be set to a large value, while it will be unavoidable in the case of circular field lines in the box. Marks of 0, 7, and 8 will return a NaN value of $Q$. For a spherical grid, a mark of 3 (or 4) can be found if 
$\vartheta_\mathrm{min}$ (or $\vartheta_\mathrm{max})$ is not very close to $-\pi/2$ (or $\pi/2$), otherwise any mark of 3 (or 4) will not appear in the output.

\begin{table}
\caption{The location types of two terminated points of a field line in FastQSL.
}
\label{tb:target}
\centering
\begin{tabular}{cc}
\toprule
Mark & Location type\\
\hline
0 & Inside the box \\
1 & $z=z_\textrm{min}$ or $r=r_\textrm{min}$ \\
2 & $z=z_\textrm{max}$ or $r=r_\textrm{max}$ \\
3 & $y=y_\textrm{min}$ or $\vartheta=\vartheta_\textrm{min}$ \\
4 & $y=y_\textrm{max}$ or $\vartheta=\vartheta_\textrm{max}$ \\
5 & $x=x_\textrm{min}$ or $\varphi=\varphi_\textrm{min}$ \\
6 & $x=x_\textrm{max}$ or $\varphi=\varphi_\textrm{max}$ \\
7 & edge or corner \\
8 & where $\bm{B} = \bm{0}$ or NaN \\
\bottomrule
\end{tabular}
\end{table}

\item $Q,\,Q_\perp$. By default, these values are defined by the coordinate mapping between two boundaries. If a field line integration is stopped before reaching the boundary (usually due to an insufficient maximum number of steps), 
the value of $Q$ becomes undefined (NaN), and $Q_\perp$ is calculated from the mapping of $\bm{U}, \bm{V}$ at the terminated points.

\item 
$Q_\mathrm{local}$. \cite{tassev2017qsl} suggested calculating $Q$ through a localized mapping to quantify the local magnetic connectivity change within a volume. In FastQSL, this value is denoted as $Q_\mathrm{local}$ and is calculated from the mapping of the local coordinates on a spherical surface centered at the launch point and having a radius of $r_\mathrm{local}$. 
If a field line is terminated at a boundary and the distance from the launch point to the terminated points is smaller than $r_\mathrm{local}$, the coordinate on this boundary is applied for the mapping. 
$Q_\mathrm{local}$ attains high values only within a distance of $r_\mathrm{local}$ from the (quasi-)separators, while its values remain low throughout the rest of the QSLs. 
FastQSL can compute $Q_\mathrm{local}$ using either Method \Rone or Method \Rtwo.
Figure \ref{fig:qlocal} provides an example of $Q_\mathrm{local}$ with the same analytical quadrupole field defined in \cite{FastQSL}:
\begin{align} 
\bm{B}_\mathrm{quadrupole}(\bm{r})= 
\sum_{i=1}^{4}\, q_i\, \frac{\bm{r}-\bm{r}_i}{|\bm{r}-\bm{r}_i|^3},
\label{eq:quadrupole}
\end{align}
where $\bm{r}_1=(-1.5,\,0,\,-0.5)$, $\bm{r}_2=(-0.5,\,0,\,-0.5)$, $\bm{r}_3=(0.5,\,0,\,-0.5)$, $\bm{r}_4=(1.5,\,0,\,-0.5)$ are the locations of 4 magnetic charges, and $\{q_1,\,q_2,\,q_3,\,q_4\}=\{1,\,-1,\,1,\,-1\}$ are the strengths of the magnetic charges.
Since Method \Rtwo will miss all of the separatrices \citep{FastQSL} for this field, we compute $Q_\mathrm{local}$ with  Method \Rone.
In the plane of $y=0$, the spatial extent of QSLs shown by $Q_\mathrm{local}$ grows with $r_\mathrm{local}$. For $r_\mathrm{local}=0.05$, the displayed structure is almost the separator. Another example of locating one segment of a separator with $Q_\mathrm{local}$ is shown in Figure \ref{fig:FMG-Quadrupole2}.
\item Length of field line.
\item
Twist number \citep{berger2006writhe}
\begin{align}
T_w=  \int_L^{}\frac{(\nabla\times\bm{B})\cdot\bm{B}}{4\,\pi\, B^2}\textrm{d}s,
\label{eq:tw}
\end{align}
where $L$ is a segment of a field line.
\item $\bm{B}$ and $\nabla\times\bm{B}$. Field-line tracing can be disabled by setting the maximum number of tracing steps to zero, allowing for rapid interpolation onto the output grid. This feature can also be used to interpolate other quantities, such as density, pressure, temperature, or to calculate the vorticity of velocity.

\end{enumerate}

\subsection{Derived Products} \label{s:derived_products}

\subsubsection{Two Parameters Used for Modeling Solar Wind Speed}
\begin{figure}[ht]
    \centering
    \includegraphics[width=\linewidth]{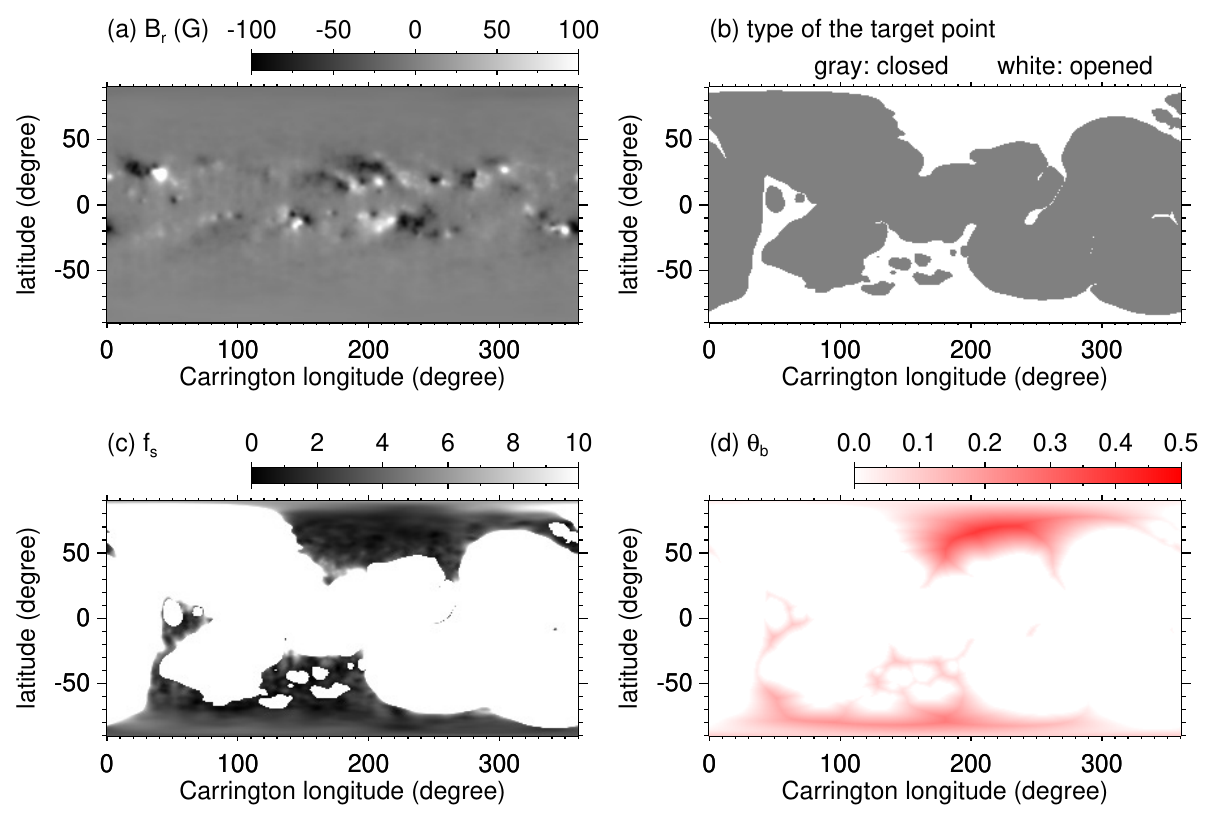}
    \caption{Parameters used for modeling solar wind speed. The magnetic field is extrapolated from the synoptic magnetogram of FMG using PFSS Model; the field of Quadrupole2 is not embedded here. $R_0=1.04~R_\textrm{sun}$, $R_1=2.5~R_\textrm{sun}$.
    (a) $B_r$ at $r=R_0$. 
    (b) Target point type.
    (c) $f_s$ at $r=R_0$. For a closed field line, its $f_s$ is set to 1000.
    (d) $\theta_b$ at $r=R_0$. For a closed field line, its $\theta_b$ is set to 0.
    }
    \label{fig:par2solarwind}
\end{figure}

\cite{2003WSAmodel} found the solar wind speed at the first Lagrangian point from December 1994 to the end of 1995 can be roughly modeled by
\begin{align}
v_\textrm{sw}=265+\dfrac{25}{f_s^{2/7}} \left(5-1.1\times \exp(1-(\theta_b/4)^2)\right)^2~\mathrm{km/s},
\end{align}
and the parameters in this formula are defined as:
\begin{enumerate}
\item Magnetic field expansion factor
\begin{align}
f_\mathrm{s}(\varphi,\,\vartheta,\,r)=
\left(\dfrac{R_0}{R_1}\right) ^2
\dfrac{B_r(\varphi_0,\,\vartheta_0,\,R_0)}
{B_r(\varphi_1,\,\vartheta_1,\,R_1)}, 
\end{align}
where $(\varphi_0,\,\vartheta_0,\,R_0)$ 
are the target coordinates traced from $(\varphi,\,\vartheta,\,r)$ to the inner boundary of $r=R_0$, and $(\varphi_1,\,\vartheta_1,\,R_1)$ are the target coordinates traced from $(\varphi,\,\vartheta,\,r)$ to the outer boundary of $r=R_1$.
\item  $\theta_b(\varphi,\,\vartheta,\,r)$, 
the minimum angular distance of an open-field footpoint from a coronal hole boundary.
\end{enumerate}
As the example in Figure \ref{fig:par2solarwind} illustrates, these
two parameters  can be derived from the coordinate mapping of FastQSL
\footnote{See \url{https://github.com/el2718/par2solarwind} for an example.}.

\subsubsection{Slip-Squashing Factors}

Slip-squashing factors $Q_\mathrm{sf}$ and $Q_\mathrm{sb}$ \citep{Titov2009} are defined by two field line mappings and two boundary flow mappings between two instants; their large values define the surfaces that border the reconnected or to-be-reconnected magnetic flux tubes for a given period of time during the magnetic evolution. 

\begin{figure}[ht]
    \centering
    \includegraphics[width=\linewidth]{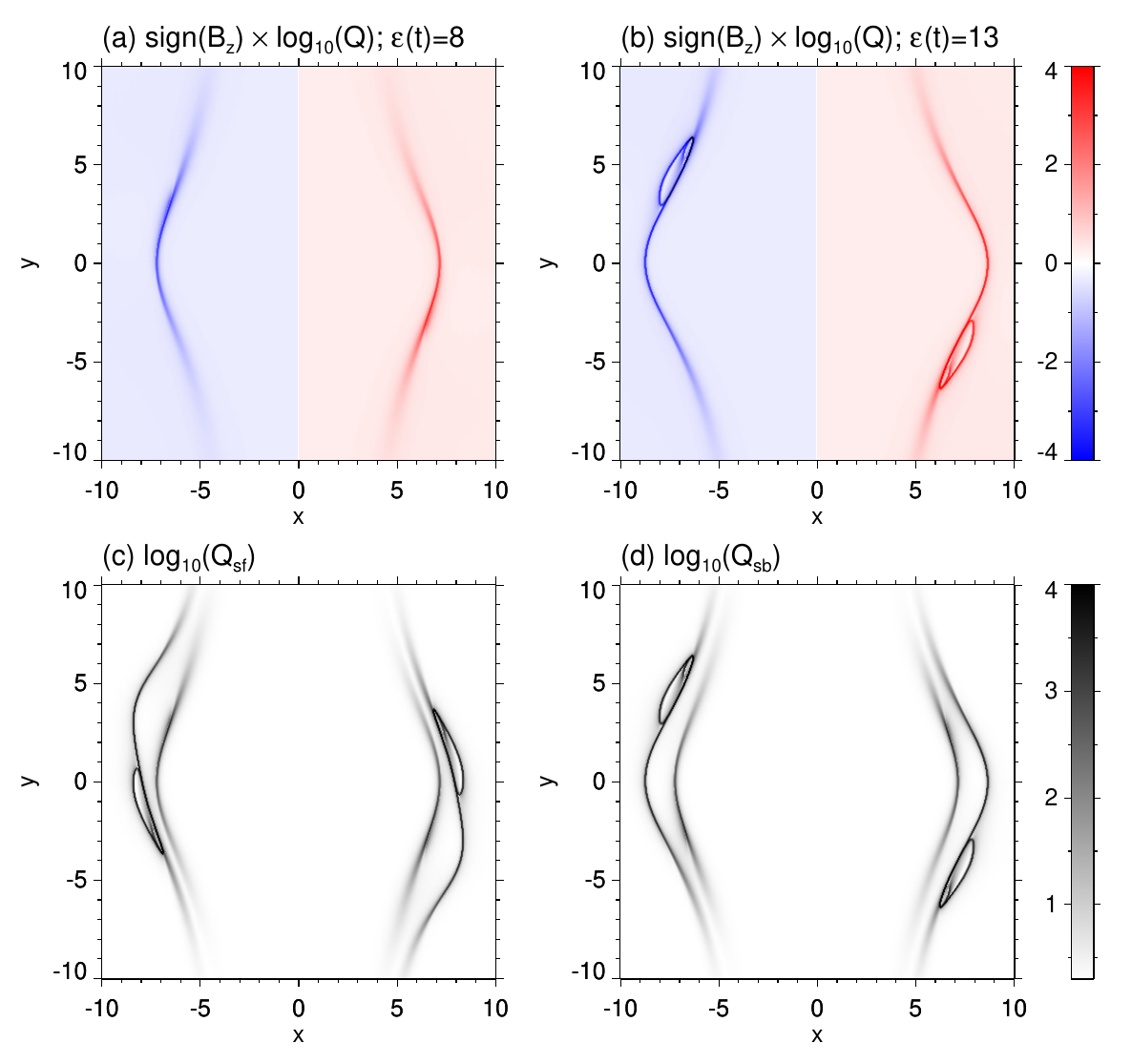}
    \caption{
    Distributions of squashing factor $Q$ and slip-squashing factors at $z=0$.
    (a-b) $Q$ at $\varepsilon(t) =8, 13$. (c-d)
    $Q_\mathrm{sf}$ and $Q_\mathrm{sb}$ between these two instants.
    }
    \label{fig:slip_plot}
\end{figure}

Up to now, no open-source code has been available to compute these quantities. 
For the case of static boundaries, one can now compute the slip-squashing factors using the coordinate mapping provided by FastQSL. Following the initial coordinate mapping within the first magnetic field, the resulting mapped coordinates can be used as a seed grid for applying FastQSL to the second magnetic field. For the field evolution of Example 1 in \cite{Titov2009},
\begin{align}
(B_x,\,B_y,\,B_z) =
\left(-1 -\frac{\varepsilon(t)\,(1-(z/5)^2)}{(1+(y/5)^2)\,(1+(z/5)^2)^2},\,0.2,\,x \right),
\end{align}
Figure \ref{fig:slip_plot} successfully reproduces the plots of Figure 6 in \cite{Titov2009}. \footnote{See \url{https://github.com/el2718/slipq} for the example.}

\section{Summary}\label{s:summary}
FastQSL has been substantially enhanced to serve most purposes of magnetic connectivity analysis. 
One significant development is the support for data on a spherical mesh, thus solving the singularity issue of tracing field lines around the polar regions by introducing a second coordinate system for the polar regions.
The transformation between the two spherical coordinate systems can also facilitate the description of data from the Solar Polar-orbit Observatory (SPO) mission \citep{Deng2025} in future studies.

The output grid of FastQSL now can have any 0-3 dimensional arbitrary shape. 
Beyond the existing capabilities of calculating twist number, the coordinates of two end points of field lines, and length of field lines in the first version, FastQSL now enables the interpolation of the magnetic field and the electric current density on the output grid, along with the ability to save the traced magnetic field lines. A (quasi-)separator, where it is easy to initiate magnetic reconnection, can be captured with high values of $Q_\mathrm{local}$.

FastQSL can also facilitate the calculation of other quantities related to magnetic connectivity. By using the coordinate mapping from FastQSL, we can accurately obtain the two parameters used for modeling solar wind speed over the whole 3-D domain.
Along with FastQSL, we also release the first open-source code for 
slip-squashing factors with static boundaries. 
This enables precise identification of reconnection sites and quantification of reconnection rates in a magnetohydrodynamic (MHD) simulation \citep{Titov2009}. 
Consequently, it becomes possible to investigate the locations and emission flux of solar flares associated with reconnection processes, even in a zero-$\beta$ MHD simulation.

The magnetic field line helicity $\oint_L \vec{A} \cdot \vec{b}\, \dd s$ is an alternative way to quantify magnetic helicity, 
where $\vec{A}$ is the vector potential of magnetic field.
\cite{Yeates2018} and \cite{Moraitis2019} derived the expression for one segment of a field line, i.e. the relative magnetic field line helicity.  
Since the path of a magnetic field line and $\vec{B}$ on the path are products of FastQSL 2, if $\vec{A}$ is also computed, the relative magnetic field line helicity can be derived.

We note that UFiT supports cylindrical coordinates. While a cylindrical grid is seldom used in practice, it significantly increases the complexity of the code. We decide to refrain from further development at this stage, and development can be resumed in the future if public needs in some aspects arise.

\acknowledgments
J.C. appreciates Suo Liu for providing the FMG data, Valentin Aslanyan for assistance with UFiT, and Ronald M. Caplan and Cooper Downs for assistance with MapFL.
This work is supported by the Strategic Priority Research Program of the Chinese Academy of Sciences (Grant No. XDB0560000), National Key R$\&$D Program of China (Grant No. 2022YFF0503003), China's Space Origins Exploration Program, the National Natural Science Foundation of China (NSFC; Grant No. 12233012, 12203102, 42274204), the mobility program (M-0068) of the Sino-German Science Center, and Deutsche Forschungsgemeinschaft (DFG, German Research Foundation) – grant 520867050.

\appendix 

\section{The Spherical Form of $Q$ Including the Coordinates on an Auxiliary Cross Section} \label{ap:method1}

In FastQSL, $S_w$ is a boundary for mapping, and $(w_1,\,w_2)$ 
are orthogonal coordinates, which are
two elements of $(\varphi,\, \vartheta,\,r)$.
Therefore for the contravariant metric tensor
\begin{align}
g_w^{11} = 1/g_{w,11},\,
g_w^{22} = 1/g_{w,22},\,
g_w^{12} = 1/g_{w,21}=g_w^{21} = 1/g_{w,12}=0.
\end{align}
For example, if $S_w$ is the surface of $r=R_\mathrm{sun}$, 
then $(w_1,\,w_2)=(\varphi,\, \vartheta)$.
Then 
$g_{w,11}=(R_\mathrm{sun}\, \cos\vartheta)^2$, $g_{w,22}=R_\mathrm{sun}^2$, 
$g_{w}^{11}=(R_\mathrm{sun}\, \cos \vartheta )^{-2}$, $g_{w}^{22}=R_\mathrm{sun}^{-2}$, 
$g_{w,12}=g_{w,21}=g_{w}^{12}=g_{w}^{21}=0$, $g_w=R_\mathrm{sun}^4\,(\cos \vartheta)^2$.
The forms on $S_u$, $S_h$ are similar.

Considering the mapping from $S_h$, we have
\begin{align}
  \pt{w_i}{u_k}&= \sum_{m}
  \pt{w_i}{h_m}\,\pt{h_m}{u_k},
\end{align}
and 
\begin{align}
\renewcommand{\arraystretch}{1.6}
\left(
\begin{array}{rr}
\pt{h_1}{u_1} & \pt{h_1}{u_2}\\[5pt]
\pt{h_2}{u_1} & \pt{h_2}{u_2}
\end{array} 
\right)
=
\left.\left(
\begin{array}{rr}
\pt{u_2}{h_2}  & -\pt{u_1}{h_2}\\[5pt]
-\pt{u_2}{h_1} & \pt{u_1}{h_1}
\end{array} 
\right)\,
\right/\,
\ptn{(u_1,\,u_2)}{(h_1,\,h_2)}.
\end{align}
Similar to Equation \eqref{eq:Delta}, we also have
\begin{align}
\sqrt{\frac{g_u}{g_h}}\,
\left|\ptn{(u_1,\,u_2)}{(h_1,\,h_2)}\right|
=\left|\frac{B_{n,h}}{B_{n,u}}\right|.
\end{align}
We denote
\begin{align}
D&=
\left(
\begin{array}{rr}
  \pt{w_1}{h_1} & \pt{w_1}{h_2} \\[8pt]
  \pt{w_2}{h_1} & \pt{w_2}{h_2}
  \end{array} 
  \right)
  \left(
\begin{array}{rr}
  \pt{u_2}{h_2} &-\pt{u_1}{h_2} \\ [8pt]
 -\pt{u_2}{h_1} & \pt{u_1}{h_1}
  \end{array} 
  \right) 
=\left(
  \begin{array}{rr}
    \pt{w_1}{h_1}\, \pt{u_2}{h_2} -\pt{w_1}{h_2}\pt{u_2}{h_1} &
  \ \ -\pt{w_1}{h_1}\, \pt{u_1}{h_2} +\pt{w_1}{h_2}\pt{u_1}{h_1} \\[8pt]
    \pt{w_2}{h_1}\, \pt{u_2}{h_2} -\pt{w_2}{h_2}\pt{u_2}{h_1} &
  \ \ -\pt{w_2}{h_1}\, \pt{u_1}{h_2} +\pt{w_2}{h_2}\pt{u_1}{h_1} \\[8pt]
    \end{array} 
    \right).
\end{align}
Finally, we obtain
\begin{align}
Q= (
g_{w,11}\,g_{u,22}\,D_{11}^2+
g_{w,11}\,g_{u,11}\,D_{12}^2+
g_{w,22}\,g_{u,22}\,D_{21}^2+
g_{w,22}\,g_{u,11}\,D_{22}^2)\,
\frac{1}{g_h}\,|B_{n,u}\,B_{n,w}/B_{n,h}^2|.
\end{align}

\section{Transformations between Two Spherical Coordinate Systems} \label{ap:transformation}
Given that $r$ is a shared coordinate of two spherical coordinate systems, the $r$-component of a vector remains unchanged. Thus, the transformation focuses solely on the components associated with the two angular coordinates.
See Figure \ref{fig:coordinates} for the geometrical relationship between the coordinates $(\varphi,\vartheta)$ and  $(\varphi_2,\vartheta_2)$.

\subsection{Coordinates}
The coordinate transformations are given by
\begin{align}
\varphi&=
\left\{
\begin{array}{ll}
  \arccos(x/\sqrt{x^2+y^2}), &\quad\text{if $y\ge0$;}\\
  2\,\pi-\arccos(x/\sqrt{x^2+y^2}), &\quad\text{if $y < 0$.}
\end{array}
\right.\\
\varphi_2&=
\left\{
  \begin{array}{ll}
    \arccos(y/\sqrt{z^2+y^2}), &\quad\text{if $z\ge0$;}\\
   2\,\pi-\arccos(y/\sqrt{z^2+y^2}), &\quad\text{if $z < 0$.}
  \end{array}
  \right.\\
\vartheta&
=\arcsin(\cos \vartheta_2\,\sin \varphi_2),\\
\vartheta_2&
=\arcsin(\cos \vartheta\,\cos \varphi).
\end{align}

\subsection{Vectors}
The corresponding unit elementary vectors of $\varphi_2, \vartheta_2$ are
\begin{align}
\ept&= -\sin \varphi_2\,\ey+ \cos \varphi_2\,\ez,\\
\ett&=\cos \vartheta_2\,\ex-\sin\vartheta_2\,\cos \varphi_2\,\ey-\sin\vartheta_2\,\sin \varphi_2\,\ez.
\end{align}

The transformations of magnetic field are given by
\begin{align}
\left(
\setlength{\arraycolsep}{2pt}
\begin{array}{l}
  B_{\varphi_2}\\
  B_{\vartheta_2}
\end{array}
\right)
&=
\left(
\begin{array}{ll}
 \ept \cdot \ep & \ept \cdot \et\\
 \ett \cdot \ep & \ett \cdot \et
\end{array}
\right)
\,
\left(
\setlength{\arraycolsep}{2pt}
\begin{array}{l}
  B_{\varphi}\\
  B_{\vartheta}
\end{array}
\right).
\label{eq:transB}
\end{align}
The expressions for $U_{\varphi_2},\, U_{\vartheta_2},\,V_{\varphi_2},\,V_{\vartheta_2},
\,(\nabla \times \bm{B})_{\varphi_2}, \,(\nabla \times \bm{B})_{\vartheta_2}$
and the corresponding reverse transformations
are similar. 

\subsection{Directional Derivatives}
The transformations of directional derivatives of coordinates are given by
\begin{align}
\renewcommand{\arraystretch}{1.3}
\dt{}{s}
\left(
\setlength{\arraycolsep}{2pt}
\begin{array}{l}
\varphi_2\\
\vartheta_2
\end{array}
\right)=
\left(
  \begin{array}{rr}
    \dfrac{\cos \vartheta}{ \cos \vartheta_2} \, \ept \cdot \ep & 
     \dfrac{1}{ \cos \vartheta_2}              \, \ept \cdot \et\\
    \cos \vartheta \, \ett \cdot \ep &  \ett \cdot \et 
  \end{array}
\right)\,
\dt{}{s}
\left(
\setlength{\arraycolsep}{2pt}
\begin{array}{l}
\varphi\\
\vartheta
\end{array}
\right).
\end{align}
The reverse transformations are similar.

For Equations \eqref{eq:dup} and \eqref{eq:dut},
\begin{align}
\dt{}{s}
\left(
\setlength{\arraycolsep}{2pt}
\begin{array}{l}
\upt\\
\utt
\end{array}
\right)
=&
\left\{
\dt{\varphi_2}{s}
\,\sin\vartheta_2
\left(
\begin{array}{rr}
  \ett \cdot \ep &  \ett\cdot \et\\
 -\ept \cdot \ep & -\ept\cdot \et
\end{array}
\right)
+ 
\dt{\varphi}{s}
\sin\vartheta
\left(
\begin{array}{rr}
  \ept \cdot \et & -\ept\cdot \ep\\
  \ett \cdot \et & -\ett\cdot \ep
\end{array}
\right)
\right\}
\left(
\setlength{\arraycolsep}{2pt}
\begin{array}{l}
\up\\
\ut
\end{array}
\right)
+ \notag\\&
\left(
\begin{array}{ll}
\ept \cdot \ep & \ept \cdot \et\\
\ett \cdot \ep & \ett \cdot \et
\end{array}
\right) \,
\dt{}{s}
\setlength{\arraycolsep}{2pt}
\left(
\begin{array}{l}
\up\\
\ut
\end{array}
\right).
\end{align}
The forms for $V_{\varphi_2},\,V_{\vartheta_2}$ 
and the corresponding reverse transformations
are similar. 

\section{Precision improvement for Terminated Points} \label{ap:boundary_trim}

We illustrate our approach in the case that a foot point should be terminated at the boundary of $z=z_\mathrm{max}$. 
In FastQSL, the main tracing integrates
\begin{align}
\frac{\mathrm{d}\,\bm{r}}{\mathrm{d}s} = \frac{\bm{B}}{B}
\label{eq:rb}
\end{align}
by 3/8-rule \texttt{RK4} or \texttt{RKF45}.
When a field line is terminated at $\bm{r}_{n+1}$ which is outside the box, and the last step inside the box is $\bm{r}_{n}$, then usually both $z_{n+1}$ and $z_{n}$ cannot be $z_\mathrm{max}$. 

In most cases, our coordinate correction at the boundaries uses 
\begin{align}
\frac{\mathrm{d}\,\bm{r}}{\mathrm{d}z} = \frac{\bm{B}}{B_z}
\label{eq:rbz}
\end{align}
to trace one step by 3/8-rule \texttt{RK4} with the step-size of $\mathrm{d}z = z_\mathrm{max}-z_{n}$ from $\bm{r}_{n}$, then the next step will be exactly located at $z=z_\mathrm{max}$. 

For some field lines that are quasi-tangent to the boundary, $B_z$ could be very small and cause a large error, then FastQSL regards the linear interpolation 
\begin{align}
\bm{r}_\mathrm{foot} = 
\frac{
(z_{n+1}-z_\mathrm{max})\,\bm{r}_{n}+
(z_\mathrm{max}-z_{n})\,\bm{r}_{n+1}}
{z_\mathrm{n+1}-z_\mathrm{n}}
\label{eq:trim2}
\end{align}
as the final foot point at the boundary.

\section{
Expression for a Cross Section Surface Perpendicular to the Photosphere} \label{ap:great_circle}

For two unit vectors $\bm{e}_{r,0}$ and  $\bm{e}_{r,1}$ on a unit sphere, 
$\bm{e}_{r,1}$
can be expressed using an anticlockwise rotation
around the axis $\bm{a}=\bm{e_{r,0}}\times\bm{e_{r,1}}/|\bm{e_{r,0}}\times\bm{e_{r,1}}|$
with the arc distance of 
$\varPhi_{01}=\arccos(\bm{e_{r,0}} \cdot \bm{e_{r,1}})$.
According to Equation (4) of \cite{rotation},
any other unit vector $\bm{e}_{r,2}$ on the great circle connecting $\bm{e}_{r,0}$ and  $\bm{e}_{r,1}$, 
which has the arc distance of $ \varPhi_{02}$ from $\bm{e}_{r,0}$, 
can be expressed as
\begin{align}
  \bm{e_{r,2}}&=
  \bm{e_{r,0}} \,\cos \varPhi_{02}+
\bm{a} \times\bm{e_{r,0}} \,\sin \varPhi_{02}.
  \label{eq:er2}
\end{align}
Then all unit vectors on the great circle can be generated by Equation \eqref{eq:er2}.
The cross section perpendicular to the photosphere can be expanded 
from the great circle in the $r$-direction.

\bibliography{cite}
\clearpage
\end{document}